\documentclass[useAMS,usenatbib,fleqn]{mnras}

\usepackage{natbib}
\usepackage{subfig}
\usepackage{graphicx}
\usepackage{etex}
\usepackage{amsmath}
\usepackage{amssymb}
\usepackage[titletoc,title]{appendix}
\usepackage{tikz}
\usetikzlibrary{shapes,arrows}
\usepackage{hyperref}
\hypersetup{colorlinks=true,linkcolor=blue,citecolor=blue,filecolor=blue,urlcolor=blue}

\usepackage{float}
\usepackage{xfrac}

\title[]{Observations of the Initial Formation and Evolution of Spiral galaxies at $1 < z < 3$ in the CANDELS fields}
\author[Margalef-Bentabol et al.]{Berta Margalef-Bentabol$^{1}$\thanks{Berta.Margalef@nottingham.ac.uk}, Christopher J. Conselice$^{2}$, Boris Haeussler$^{3}$, \newauthor Kevin Casteels$^{4}$, Chris Lintott$^{5}$, 	Karen Masters$^{6}$, Brooke Simmons$^{7}$\\
$^{1}$Department of Physics and Astronomy, University of Pennsylvania, Philadelphia, PA 19104, USA\\
$^{2}$Jodrell Bank Centre for Astrophysics, University of Manchester, Oxford Road, Manchester UK \\
$^{3}$European Southern Observatory, Alonso de Cordova 3107, Vitacura, Casilla 19001, Santiago, Chile\\
$^{4}$Department of Physics and Astronomy, University of Victoria, Victoria, BC V8P 1A1, Canada\\
$^{5}$Oxford Astrophysics, Denys Wilkinson Building, Keble Road, Oxford OX1 3RH, UK\\
$^{6}$Haverford College, 370 Lancaster Ave., Haverford, PA 19041, USA\\
$^{7}$Physics Department, Lancaster University, Lancaster LA1 4YB, UK}

\date{Accepted XXX. Received YYY; in original form ZZZ}

\pubyear{2021}

\begin{document}
\label{firstpage}
\pagerange{\pageref{firstpage}--\pageref{lastpage}}
\maketitle

\begin{abstract}

Many aspects concerning the formation of spiral and disc galaxies remain unresolved, despite their discovery and detailed study over the past $150$ years.  As such, we present the results of an observational search for proto-spiral galaxies and their earliest formation, including the discovery of a significant population of spiral-like and clumpy galaxies at $z>1$ in deep \textit{Hubble Space Telescope} CANDELS imaging. We carry out a detailed analysis of this population, characterising their number density evolution, masses, star formation rates and sizes. Overall, we find a surprisingly high overall number density of massive $M_{*} >10^{10}\mathrm{M}_{\odot}$ spiral-like galaxies (including clumpy spirals) at $z > 1$ of $0.18\,{\rm per}\, \mathrm{arcmin}^{-2}$. We measure and characterise the decline in the number of these systems at higher redshift using simulations to correct for redshift effects in identifications, finding that the true fraction of spiral-like galaxies grows at lower redshifts as $\sim$ $(1+z)^{-1.1}$. This is such that the absolute numbers of spirals increases by a factor of $\sim 10$ between $z = 2.5$ and $z = 0.5$. We also demonstrate that these spiral-like systems have large sizes at $z>2$, and high star formation rates, above the main-sequence, These galaxies represent a major mode of galaxy formation in the early universe, perhaps driven by the spiral structure itself.  We finally discuss the origin of these systems, including their likely formation through gas accretion and minor mergers, but conclude that major mergers are an unlikely cause.

\end{abstract}


\section{Introduction}\label{sec.introduction}

Spiral structures are present in many disc galaxies in the local Universe with different morphologies, ranging from grand-design, in which prominent and well-defined spiral arms can be traced over large parts of the disc, to those with a flocculent pattern, with fragmented spiral arms that have a more limited spacial extent \citep{Elmegreen11, Elmegreen14}. Since these spirals/discs make up the majority of the galaxy population in today's universe \citep[e.g.,][]{Conselice06}, understanding their development is critical for any full picture of galaxy formation. 

Different mechanisms are believed to be responsible for creating the morphology of these spiral galaxies \citep{Elmegreen90}. The formation of the grand design spirals can be explained by the so-called density wave theory \citep{Lin64, Bertin89}. This theory suggests that the spiral pattern is a wave that moves through a spiral's disc causing its stars and gas to clump up along the wave, moving with a velocity that is independent of the rotation of the disc. The flocculent arms are believed to be produced by local instabilities \citep{Toomre90, Sellwood91}. Other studies also suggest that spiral structure can be driven by tidal effects due to an interaction with a companion galaxy, or the presence of a central bar \citep{Dobbs10, Hart18}. However, there is no clear answer to how the different spiral patterns are produced or when they form. By examining these systems in the distant universe we may be able to determine their formation history in an empirical way that can inform the physics for their development and long lasting nature.

In the local universe, spiral galaxies are relatively common, and comprise one of the main broad categories of the Hubble sequence \citep{Sandage61}. However, they are very rare at high redshift, and only two grand-design-like spiral galaxies at $z>2$ have been confirmed spectroscopically: HDFX 28 at $z=2.011$ (\citealt{Dawson03}, which may instead be a major merger resembling a spiral) and Q2343-BX442 at $z=2.18$ \citep{Law12}, showing that such structures can be detected at high redshift, although Q2343-BX442 is relatively massive ($\log M_{\ast}=10.78$) and bright ($\textit{H}=22.04\ \mathrm{AB}$). Therefore, the current limit on instrumentation could be responsible for the low number of spiral galaxies detected at high redshift compared to lower redshifts ($z<2$). 

However, the scarcity of spirals galaxies at $z>2$ could have a physical explanation as well, since at $z>2$ discs are dynamically hot \citep{Genzel06,Law09}, which produces more often clumpy structures than spirals arms \citep{Conselice04,Elmegreen05a}. Minor mergers could also be responsible for the design of spiral patterns, as suggested by numerical simulations \citep{Bottema03, Dobbs10a}. This may explain why spiral galaxies are relatively rare, as they need their host galaxy to be massive enough to stabilise their disc, and also at the same time the presence of a minor merger whose perturbation induces the formation of a spiral pattern.

A fundamental question in galaxy evolution is therefore answering when modern spirals appear and how this occurs. As stated earlier, these objects are quite common in the local universe, but can be difficult to study in the distant universe partially due to redshift effects producing lower resolution and lower S/N image. It has been found that at high redshift ($z\sim 2$) star-forming galaxies become increasingly irregular and clumpy \citep{Elmegreen05b,Elmegreen09}. These clumpy galaxies may transition into spirals, after a gradual dispersal of the clumps to form discs \citep{Elmegreen09}. It is also still debated what impact spiral-arms have on the star formation of their discs, and whether the number of spiral arms have an effect on the star-formation \citep{Hart17}. Spirals may enhance the star formation, as they are the site of high density of gas and young stars, at least in the local universe \citep{Engargiola03, Calzetti05}. However, there are some studies which suggest that the spiral arms do not affect the overall star formation in the host galaxy \citep[e.g.][]{Elmegreen02, Moore12}. There are therefore many questions that are unanswered about the formation of disc galaxies. All of these questions can be addressed by examining how galaxies with spirals evolve through cosmic time, yet this has never been achieved before on a statistically large sample.

At high redshift at about $z>2$ the number density of galaxies is dominated by peculiar systems \citep{Mortlock15}. However, an important population of disc-dominated massive galaxies has also been found\citep{Bruce12, Margalef16}. As redshift decreases galaxies experience a change in morphology and start to transition from peculiar to Hubble type systems \citep{Conselice05,Mortlock15,Huertas16}. This evolution is such that by $z=1$ galaxies formed by a bulge and a disc start to dominate the galaxy population \citep{Margalef16}. The majority of these high redshift disc-dominated galaxies are actively forming stars, growing is size as redshift decreases \citep{Margalef18,Whitney19}. 

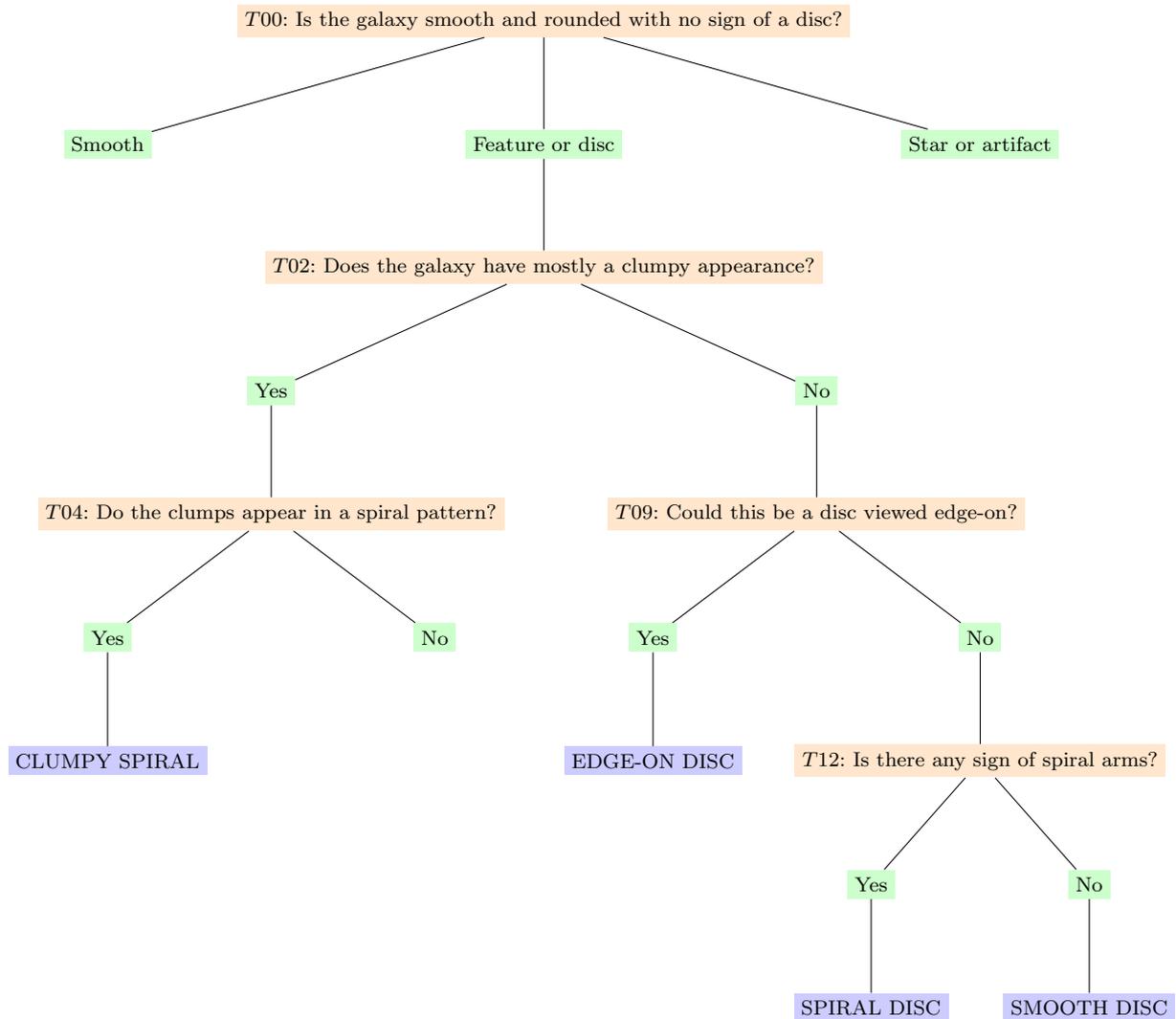
\begin{figure*}
\centerline{\begin{tikzpicture}[auto]
\tikzstyle{level 1}=[sibling distance=60mm,level distance=14ex] 
\tikzstyle{level 2}=[sibling distance=55mm,level distance=14ex] 
\tikzstyle{level 3}=[sibling distance=75mm,level distance=14ex] 
\tikzstyle{level 4}=[sibling distance=25mm,level distance=14ex] 
\tikzstyle{level 5}=[sibling distance=45mm,level distance=14ex] 
\tikzstyle{level 6}=[sibling distance=15mm,level distance=14ex] 
\tikzstyle{level 7}=[sibling distance=30mm,level distance=14ex] 
\node [rectangle,fill=orange!20] (t00) {\footnotesize{$T00$: Is the galaxy smooth and rounded with no sign of a disc?}}
child{  node [rectangle,fill=green!20] (smooth) {\footnotesize{Smooth}}  } 
child{  node [rectangle,fill=green!20] (feature) {\footnotesize{Feature or disc}}
     child{  node [rectangle,fill=orange!20] (t02) {\footnotesize{$T02$: Does the galaxy have mostly a clumpy appearance?}}
          child{  node [rectangle,fill=green!20] (t02yes) {\footnotesize{Yes}}
               child{  node [rectangle,fill=orange!20] (t04) {\footnotesize{$T04$: Do the clumps appear in a spiral pattern?}}
                    child{  node [rectangle,fill=green!20] (t04yes) {\footnotesize{Yes}}
                         child{  node [rectangle,fill=blue!20] (clumpy) {\footnotesize{CLUMPY SPIRAL}}  }  }
                    child{  node [rectangle,fill=green!20] (t04no) {\footnotesize{No}}  }  }  }
          child{  node [rectangle,fill=green!20] (t02no) {No}
               child{  node [rectangle,fill=orange!20] (t09) {\footnotesize{$T09$: Could this be a disc viewed edge-on?}}
                    child{  node [rectangle,fill=green!20] (t09yes) {\footnotesize{Yes}}
                         child{  node [rectangle,fill=blue!20] (edge) {\footnotesize{EDGE-ON DISC}}  }  }
                    child{  node [rectangle,fill=green!20] (t09no) {\footnotesize{No}}
                         child{  node [rectangle,fill=orange!20] (t12) {\footnotesize{$T12$: Is there any sign of spiral arms?}}
                              child{  node [rectangle,fill=green!20] (t12yes) {\footnotesize{Yes}}
                                   child{  node [rectangle,fill=blue!20] (spiral) {\footnotesize{SPIRAL DISC}}}  }
                              child{  node [rectangle,fill=green!20] (t12no) {\footnotesize{No}}
                                   child{  node [rectangle,fill=blue!20] (disc) {\footnotesize{SMOOTH DISC}}  }  }  }  }  }  }  }  }
child{  node [rectangle,fill=green!20] (star) {\footnotesize{Star or artifact}}  };
\end{tikzpicture}}
\caption[The decision tree for Galaxy Zoo CANDELS]{Simplified version of the decision tree of Galaxy Zoo CANDELS from which we obtain our morphological classifications. This includes the questions that leads to one of the classifications of interest for this work (blue boxes): spiral disc, clumpy spirals, smooth discs and edge-on discs.} \label{fig1}
\end{figure*}

The most star-forming disc-dominated galaxies also appear to be rounder and clumpier compared to other disc galaxies \citep{Bruce14a}. There is evidence that stellar mass plays an important role in the evolution of these high redshift galaxies, with the most massive galaxies becoming preferentially systems with a bulge and a disc earlier \citep{Margalef16}. However, it is still unclear what role spiral structure plays in these high redshift massive galaxies and how they are forming.

In this Paper we investigate the properties of spiral discs and their evolution through cosmic time and especially from their formation at $z > 1.5$. We compare the evolution of smooth discs without spiral arms for comparison purposes, although this has to be carefully considered due to resolution effects. Overall our goal is to determine how the first disc galaxies formed in an observational way and to compare with models and other galaxy types.

\begin{figure*}
  \centering
  \includegraphics[width=1\linewidth]{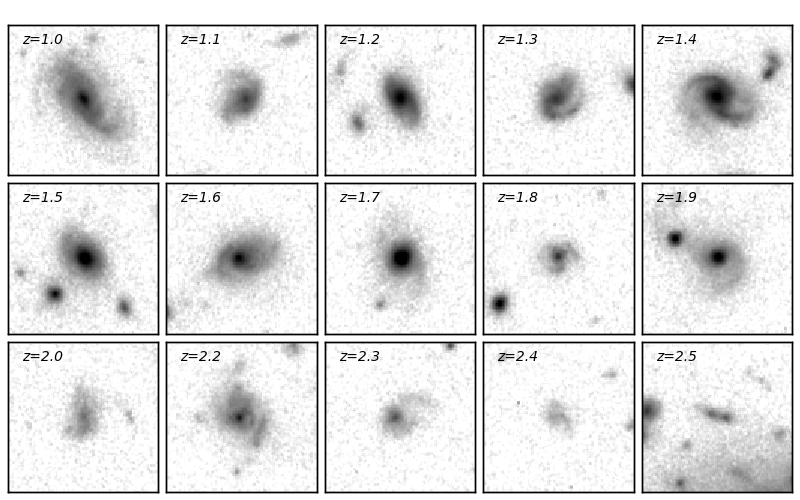}
  \caption{Examples of galaxies classified as spiral discs in Galaxy Zoo at a variety of redshifts. On the top right of each panel is the redshift of the galaxy imaged. Spiral-like structure can be seen in most of these galaxies. We discuss the possibility later in this paper that some of these are not actual spiral types, and the biases in classification that may exist. The postage stamps are $6\times 6$ arcseconds in size.}\label{fig2}
\end{figure*}

The structure of this Paper is as follow. In Section \ref{sec.data} we describe the data used in this work. Section \ref{sec.method} is devoted to describing how we select the different types of galaxies for our study. In Section \ref{sec.results} the main results of the Paper are gathered, and in Section \ref{sec.conclusions} we discuss and summarize the results. Throughout this paper we use AB magnitude units and assume the following cosmology: $H_0=70\mathrm{\ Km s}^{-1}\mathrm{ Mpc}^{-1}$, $\Omega_{\lambda}=0.7$, and $\Omega_m=0.3$.

\begin{figure*}
  \centering
  \includegraphics[width=1\linewidth]{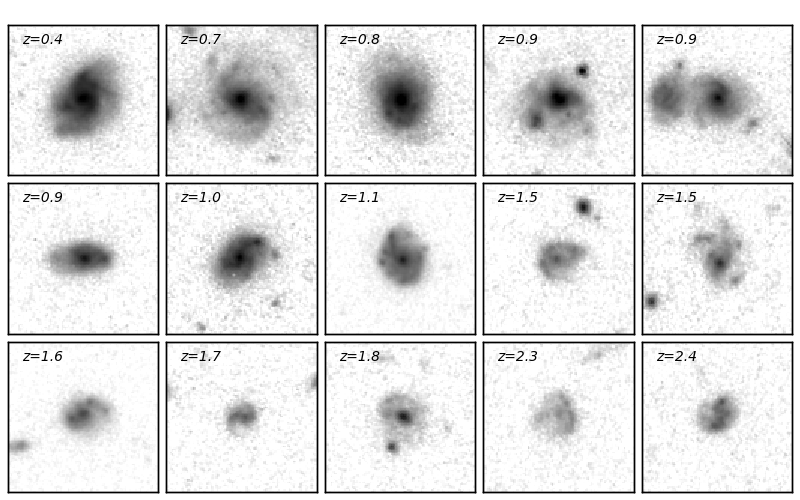}
  \caption{Examples of galaxies classified as clumpy spirals in the Galaxy Zoo methodology. On the top right of each panel we give the redshift of the galaxy imaged. The postage stamps are $6\times 6$ arcseconds in size.}\label{fig3}
\end{figure*}

\begin{figure*}
  \centering
  \includegraphics[width=1\linewidth]{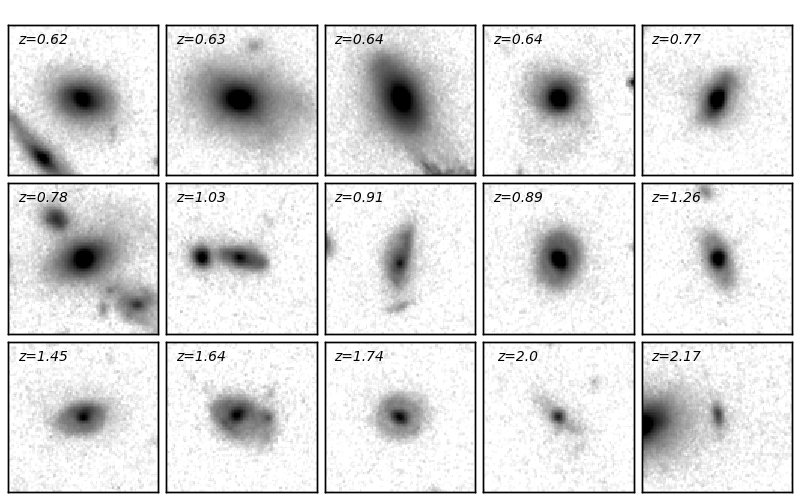}
  \caption{Examples of galaxies classified as smooth discs in the Galaxy Zoo methodology. On the top right of each panel we give the redshift of the galaxy imaged. The postage stamps are $6\times 6$ arcseconds in size.}\label{fig4}
\end{figure*}

\section{Data}\label{sec.data}
\subsection{CANDELS}

The Cosmic Assembly Near-infrared Extragalactic Legacy Survey (CANDELS; \cite{Grogin11}, \cite{Koekemoer11}) is HST Multi Cycle Treasury Program which combines optical and infrared imaging from the Wide Field Camera (WFC3) and the Advanced Camera for Surveys (ACS) to image the distant Universe. CANDELS targets five well studied fields (GOODS-S GOODS-N, UDS, EGS and COSMOS) at two distinct depths (`deep' and `wide'). The deep portion of the survey covers GOODS-S and GOODS-N. The wide portion images all CANDELS fields. In total, CANDELS covers an area of $800$ $\mathrm{arcmin}^2$.

For this paper, we use a sample of galaxies from the COSMOS, GOODS-S, and UDS fields for which classifications from the Galaxy Zoo project are available. The CANDELS fields have been imaged by a large number of multiwavelength surveys. For a detailed discussion of the CANDELS UDS photometry see \cite{Galametz13}. The multi-band photometry catalog of GOODS-S is described by \cite{Guo13} and for COSMOS by \cite{Nayyeri17}. For other and future uses of our catalog see \cite{Margalef16, Margalef18} where the data set we use is described in more detail.

\subsection{Galaxy Zoo}

The galaxy classifications that we use throughout this paper to identify spiral galaxies of various types comes from the Galaxy Zoo citizen science project. Galaxy Zoo is an online citizen science project that started in July $2007$, with a data set of around a million galaxies from the Sloan Digital Sky Survey (SDSS), in which volunteers were asked to classify galaxies into ellipticals, mergers and spirals \citep{Lintott08}. In later phases, more detail morphological classifications were achieved, by asking volunteers more complex question such as the number of spiral arms or the size of the galaxy bulges. For this work we use the classifications collected during the fourth release of Galaxy Zoo \citep{Simmons17}. This $4$th phase of Galaxy Zoo includes all detections with $H\leqslant25.5$ from COSMOS, GOODS-S and UDS in the CANDELS Survey, comprising $49\,555$ images, and it provides detailed quantitative visual morphologies, including the overall galaxy's appearance and more specific features (such as the presence of spiral arms, a clumpy appearance or prominence of the bulge). The images shown for classification to the `citizen scientist' classifiers are colour composites of the ACS-\textit{i}, WFC3-\textit{J} and WFC3-\textit{H} filters in the blue, green and red channels, respectively.

\subsection{Redshifts and stellar masses}\label{sec.data.redshifts}

We use the official catalogues of redshifts (spectroscopic redshifts are used when available, photometric redshifts otherwise) and stellar masses from CANDELS in order to have consistent measurements within the different fields. For the CANDELS UDS and GOODS-S fields, the redshifts are published in \cite{Dahlen13}, and the masses are described in \cite{Santini14}. Redshifts and stellar masses for COSMOS are found in \cite{Nayyeri17}. The redshifts and stellar masses used in this work are thus measured in a systematic way, and we check, using distributions, that there are no systematic differences between the three fields. In \cite{Santini14} dust extinctions used are range between $0 <$ E(B-V) $< 1.1$ with a Calzetti or SMC extinction curve. In \cite{Nayyeri17} the extinction values range from 0 to 0.5 mag with an extinction curve derived from the modified dust grain size distribution from \cite{Draine84}.

Photometric redshifts are estimated using the multi-wavelength photometry catalogs and adopting a Bayesian approach described in \cite{Dahlen13}, which combines the posterior redshift probability distribution from several independent SED fitting codes, thus improving the precision and reducing the number of catastrophic outliers. The typical dispersion of $z_\mathrm{photo.}$ vs. $z_\mathrm{spec.}$ spans from $0.25$ to $0.31$ in $\delta z /(1 + z)$ for the photometric redshifts.

For UDS and GOODS-S fields, stellar masses are estimated by fitting the photometry with a library of stellar synthetic SEDs from the stellar population models of \cite{Bruzual03}. Star formation histories are parametrized as exponentially declining laws, and a Chabier IMF is used. Redshifts are fixed to the photometric or spectroscopic when available. Stellar masses for the COSMOS field are calculated as the median stellar mass reported by different SED fitting methods, with the redshift fix to the photometric redshifts or spectroscopic when available. For further details of the stellar mass fitting see the original CANDELS team papers where these are described in detail \cite{Santini14, Nayyeri17}.

\subsection{Structural parameters}\label{sec.data.structural}

We use \textsc{galfitm} \citep{Haeussler13, Vika13,Vika14} and \textsc{galapagos-2} \citep{Haeussler13} to derive the structural parameters for our galaxy sample. We perform single S\'ersic profile fitting to the surface brightness of each galaxy as well as $2$-component fitting (bulge/disc decomposition). Throughout the paper we use the S\'ersic index \citep{Sersic}, effective radius and magnitudes obtained with the single S\'ersic fitting, and the bulge to total ratios from the two component fitting. However, it is worth noting that bulge to total ratios are hard to measure and are not as reliable as the other properties for single systems. Despite this in aggregate, a combination or average ratio should be reliable.

\textsc{galapagos-2} \citep[an extended version of \textsc{galapagos},][]{Barden12} uses \textsc{SExtractor} \citep{Bertin96} and \textsc{galfitm} \citep[an extended version of \textsc{Galfit3},][]{Peng02,Peng10} to perform consistent S\'ersic profile fitting to multi-band data. It provides a sophisticated wrapper around \textsc{galfitm}, that handles the fitting process from the input image to the output catalogue. \textsc{galfitm} allows each parameter in the S\'ersic profile to vary as a function of wavelength (with the user choosing the degree of wavelength dependence). Fitting a surface brightness profiles in multiple wavebands increases the accuracy and stability of each measurement (Haeussler in prep.).

For the single S\'ersic fit, we run \textsc{galapagos-2} on all three of the CANDLES fields used in this work (COSMOS, UDS and GOOD-S). A detail explanation on how the code works can be found in \cite{Barden08}. In summary, \textsc{galapagos-2} code creates postage stamps for each galaxy, and masks which are used during the fitting process after the code decides whether a neighbouring object is masked or fit simultaneously, depending on its brightness and distance. The code then calculates the sky value, that is fixed later during the fit. Finally, it prepares the input files for \textsc{galfitm} and performs the fitting for all objects. In this case, galaxy magnitudes are allowed to vary freely in each wavelength, while the S\'ersic index and effective radius are modelled as quadratic functions of wavelength in the single S\'ersic fits used here. The effective radius is constrained to be between 0.3 and 800 pixels to ensure physical meaning. The S\'ersic index within our fits ranges from $n = 0.2$ to $12$, although most objects fall between $n \sim 1-5$. Position coordinates, axis ratio and position angle are not permitted to vary with wavelength and instead ensure that the galaxy models are consistently centered.

For the B/D decomposition, \textsc{galapagos-2} uses the same postage stamps, masks, PSFs and wavelengths as in the single-S\'ersic fit, but creates new input files for \textsc{galfitm}. The main change is that the target galaxy is now fit with two S\'ersic profiles instead of one: an exponential profile (S\'ersic profile with $n=1$) and a S\'ersic profile with a free fitted $n$, to represent a disc and a bulge, respectively. The starting values are chosen according to the single-S\'ersic results (Haeussler in prep.).

\section{Method}\label{sec.method}

The focus of this work is studying the properties of spiral disc galaxies and to compare these with discs that do not show any spiral pattern, as well as to compare with other galaxy types at similar redshifts. We use the Galaxy Zoo classification scheme and catalog to construct our primary sample of galaxies, with the desired morphologies: smooth discs and spiral galaxies. The latter class of `spirals' comprises both regular spiral galaxies with either a grand-design spiral or a flocculent pattern, or clumpy spirals where several clumps appear in a spiral pattern, as oppose to as in a cluster, straight line or chain. We also include in our sample, as a separate class, edge-on discs, as for these galaxies it is not possible to know from a visual classification whether there is any spiral pattern, and therefore they are most likely a combination of the above morphologies. 

Galaxy Zoo uses a tree based structure for collecting information on morphological features. The CANDELS decision tree first asks the classifier to choose between the broad categories of smooth and rounded, feature or disc, and star or artifact ($T00$ in Figure \ref{fig1}). If the classifier has indicated in the first task that the galaxy has features or a disc, then there are follow up questions about more complex features (clumpy, edge-on viewed or spiral pattern). In Figure \ref{fig1} we show part of the Galaxy Zoo decision tree \citep[see][for a whole explanation on the decision tree]{Simmons17}, in particular the branches that lead to the classifications relevant for this work: spiral galaxies, clumpy spirals, smooth discs and edge-on discs. For a particular question, each possible answer has a vote-fraction $f$ assigned to it, and the sum of the vote fraction of all possibles answers to one question adds to one; for example, if we consider $T00$, $f_{smooth}+f_{features}+f_{artifact}=1$. To reduce the influence of less reliable classifiers, \cite{Simmons17} apply a weighting method to obtain the final vote-fractions $f$. For this work we use these weighted vote fractions.

\begin{figure*}
  \centering
  \includegraphics[width=0.33\linewidth]{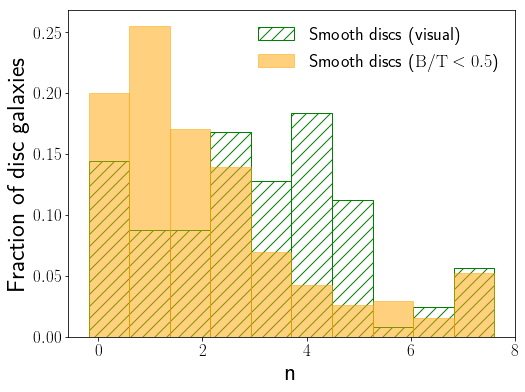}
  \includegraphics[width=0.33\linewidth]{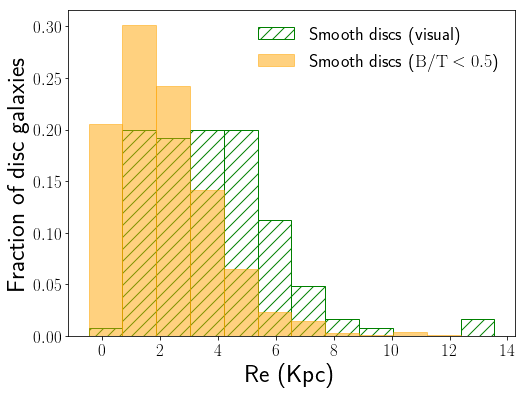}
  \includegraphics[width=0.33\linewidth]{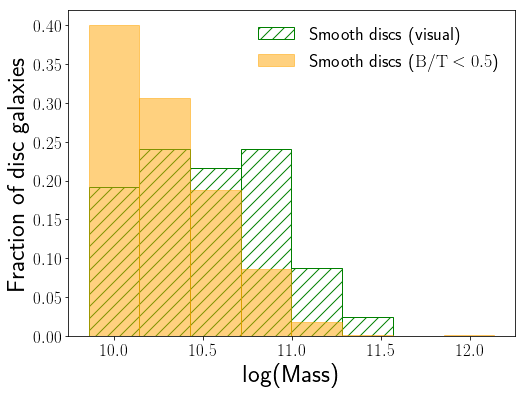}
  \caption{Comparison between smooth discs selected visually (stripped lines green) and smooth galaxies with $B/T<0.5$ (solid yellow), for galaxies with $z < 3$ and $\log M_{\ast} >10$. The left panel shows the distribution of S\'ersic index $n$, the middle, the distribution of effective radius $R_e$ and the bottom panel shows the distribution of stellar mass.}\label{fig5}
\end{figure*}

To select a sample of galaxies with a specific morphological class we apply some thresholds to the vote fractions for each question that leads to that class in turn. The thresholds used in this paper are slightly different to those proposed on \cite{Simmons17}, to allow for a more complete sample for spirals and clumpy spirals. We choose these thresholds after visually inspecting a sub-sample of galaxies, to ensure a desired level of classification completeness for each class. For example, the sample of spiral discs is comprised of galaxies for which at least $30$ per cent of the classifiers choose `Feature or disc' to answer the question T00 ($f_{features} >0.3$) and at least 50 per cent of voters answered `No' to question T02 i.e. not having clumpy appearance ($f_{clumpy}<0.5$), this can also be expressed as less than 50 per cent of voters answering `Yes' to question T02 and are classified as not being edge-on by at least 50 per cent of voters ($f_{edge-on}<0.5$) and finally being classified as `Spiral discs' by more that 50 per cent of classifiers answering question T12 ($f_{spiral}>0.5$). We summarize which criteria must be satisfied for each different type of disc used in this study to be included in our sample (blue boxes of Figure \ref{fig1}) as follow:

\begin{align*}
\intertext{$\bullet$ Spiral discs:}
f_{features} >0.3\ \wedge\ f_{clumpy}<0.5\ \wedge\ f_{edge-on}<0.5\ \\ \wedge\ f_{spiral}>0.5 
\intertext{$\bullet$ Clumpy spirals:}
f_{features} >0.3\ \wedge\ f_{clumpy}>0.5\ \wedge\ f_{clumpy-spiral}>0.5
\intertext{$\bullet$ Smooth discs:}
f_{features} >0.3\ \wedge\ f_{clumpy}<0.5\ \wedge\ f_{edge-on}<0.5\ \\ \wedge\ f_{spiral}<0.5
\intertext{$\bullet$ Edge-on discs:}
f_{features} >0.3\ \wedge\ f_{clumpy}<0.5\ \wedge\ f_{edge-on}>0.5
\end{align*}

We also require that $f_{artifact}<0.3$ and that the number of classifiers in the final question for each disc type is at least $10$. Finally, we limit our sample to galaxies with $\log M_{\ast}>10$, to study galaxies in the same stellar mass range as in \cite{Margalef16} and \cite{Margalef18}. Figures \ref{fig2}, \ref{fig3} and \ref{fig4} show some of the galaxies that are classified by our Galaxy Zoo classification as spirals, clumpy spirals and smooth discs, respectively. Galaxies classified as smooth discs appear to be discs galaxies, but features such as spiral arms are either not present or not visible. This selection criteria leaves us with a sample of $849$ galaxies in these for classes in total, across all of our fields.

We have to consider the fact as well that some of the galaxies we classify as one type by eye may not in fact be their real morphology due to resolution effects. For example, \cite{Simmons17} discuss the fact that some of the galaxies classified as smooth in the first question may be in fact discs without any distinctive feature, as visually it can be difficult to distinguish an elliptical galaxy from some types of smooth discs. \cite{Simmons17} propose to select those smooth discs by looking at the relative bulge and disc strength, i.e.,\ using a cut in the bulge to total fraction ($B/T < 0.5$) from a bulge to disc decomposition of the surface brightness distribution (in the \textit{H}-band). Bulge to disc decomposition has been performed within the UDS and COSMOS CANDELS fields using Galapagos-2 \citep{Haeussler13}. By adding these smooth discs to the visually classified smooth discs we have a more complete sample and one in which we can fully explore possible intrinsic structures. It is worth noting that $B/T$ ratios are not very reliable for galaxies in which the bulge is much fainter than the disc (Haeussler in prep.). In such case, GALAPAGOS is not able to properly fit the bulge, and instead the bulge profile will likely become the same as the disc, and therefore both components will have $n \sim 1$ and similar sizes. When this happens, the B/T becomes unconstrained and this may results in our sample of smooth discs with $B/T< 0.5$ being contaminated with bulge-dominated objects, as well as the sample not being complete. We quantify the possible contamination by looking at the number of galaxies in that sample that have the S\'ersic index and effective radius of the bulge within $10$ per cent of those of the disc, as this can be an indication of the B/T being unconstrained. We found that this is true for only $3$ per cent of the sample. Furthermore, we look at the B/T of our sample in all the other bands available (\textit{J, V, i}) and find that only $5$ per cent of the galaxies have a B/T in the \textit{H}-band that differs more than $10$ per cent in the other bands (if the B/T was unconstrained the B/T would differ significantly between bands). Therefore, we expect that our smooth disc sample, as selected by the B/T may have a contamination from bulge dominated galaxies of less than $8$ per cent.

\begin{figure}
  \centering
  \includegraphics[width=1\linewidth]{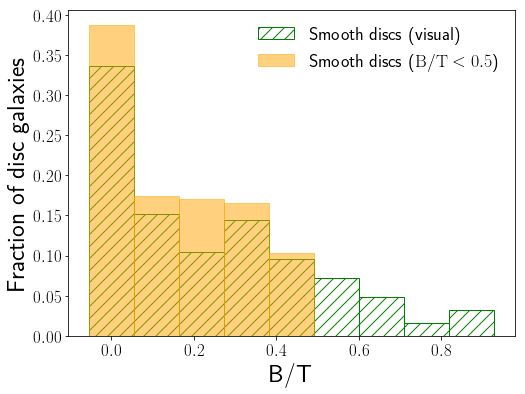}
  \caption{Distribution of $B/T$ values for smooth discs selected visually (stripped lines green) and smooth galaxies selected with $B/T<0.5$ (solid yellow). As can be seen, the distributions are fairly similar except at the highest B/T ratios where we find that the smooth discs contain a larger B/T ratio than any of the smooth discs selected by eye. This fraction is however small at 15 per cent (see text).}\label{fig6}
\end{figure}

In Figure \ref{fig5} we compare the properties of the smooth discs selected visually to those determined by using a cut in $B/T$. Visually smooth discs are larger in size and are more concentrated (higher S\'ersic index, $n$), and they have on average higher stellar masses. In Figure \ref{fig6} we compare the $B/T$ of the smooth discs selected visually and those chosen by a cut in $B/T$. As expected, the majority of the visually classified smooth discs also have $B/T<0.5$, although there is a small fraction ($15$ per cent) that have $B/T>0.5$. It is possible that some of these galaxies are misclassifications from elliptical galaxies, but the error in this is likely very small, probably close to the high B/T ratio fraction of $15$ per cent. Therefore we conclude that this is not a strong effect that will bias our results. We investigate whether there is a correlation between B/T ratio and stellar mass and find that this correlation exist particularly for smooth discs, such that of those classified as discs with B/T $> 0.5$, $60$ per cent have stellar masses log M$_{*} >10.5$ and $13$ per cent are with stellar masses log M$_{*} > 11$, but we find no correlation for the clumpy discs.

For comparison to the spirals, in the next section we use the visually classified smooth discs for our analysis. However, the main results for the spirals and clumpy spirals remain the same even when including the smooth discs selected by their $B/T$ ratios in our results. Therefore, in what follows we cannot be sure that any given one high-z disc-like galaxies is truly a disc, but as we consider trends and averages this slight contamination will not adversely affect our results.

\subsection{CAS parameters}

The analysis of the concentrations (C), asymmetries (A) and clumpiness (S) is useful to investigate the structures of galaxies. Those parameters, known as CAS parameters, are computed in \cite{Mortlock15} using the \textit{H}-band image in the CANDELS UDS field. Note that all the galaxies we measure are cleaned of nearby neighboring galaxies and have their CAS parameters measured with careful attention to centring and background correction, as described in \citep{Conselice03b}.

The \textbf{asymmetry} is found by subtracting a $180$ degrees rotated image from its centre of the galaxy to the original image plus a background subtraction \citep{Conselice00, Conselice03b}:

\begin{equation}
 A=\min\left(\frac{\sum|I_0-I_{180}|}{\sum I_0}\right)-\min\left(\frac{\sum|B_0-B_{180}|}{\sum I_0}\right),
\end{equation}

\noindent where $I_0$ is the original image and $I_{180}$ is the image after rotating it by $180$ degrees. $B_0$ is the original background image and $B_{180}$ is the background image rotated by $180$ degrees. The sum is performed over all the pixels within a matching region of the original and rotated images. The minimization on each term is done to locate the centre of the galaxy that minimize the asymmetry. See \cite{Conselice00} for more information on the details of the calculation.

\begin{figure}
	\centering
	\includegraphics[width=1\linewidth]{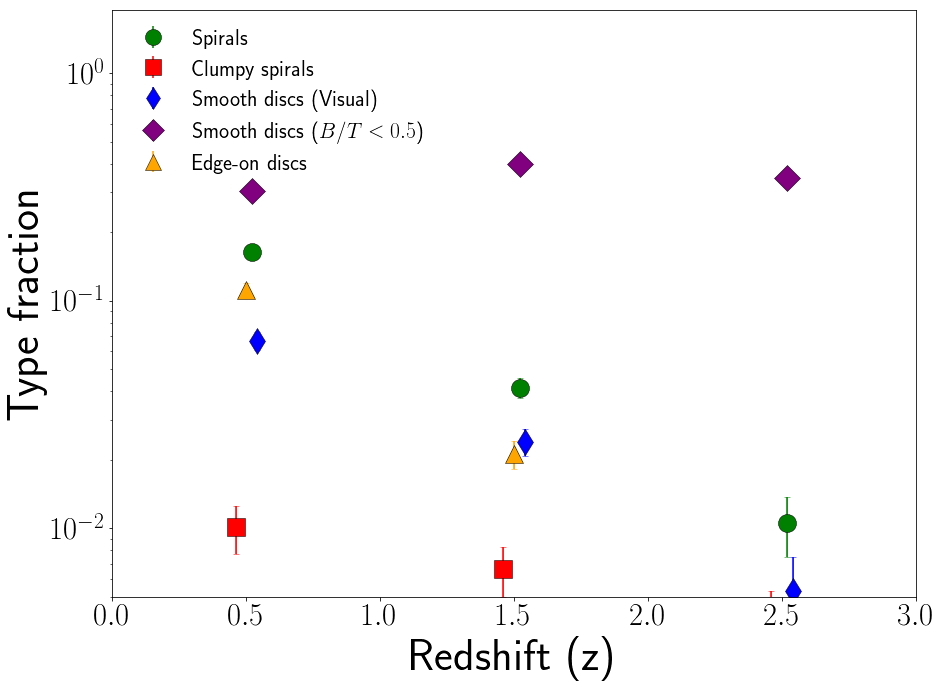}
	\caption{Fraction of different types of discs as a function of redshift without taking into account redshift effects, which we consider later in \S 4.1. Green circles are disc galaxies with an observed spiral pattern, red squares are discs with a clumpy spiral structure, yellow triangles are edge-on discs, blue diamonds are discs with no spiral pattern and purple diamonds are smooth galaxies with $B/T < 0.5$.}\label{fig7}
\end{figure}

The \textbf{concentration} parameter measures how concentrated the light in a central region is compared to a larger, less concentrated region. It strongly correlates with the S\'ersic index $n$ and the bulge to total ratio, which are both measures of the light concentration in a galaxy. The formula for computing the concentration is given by:

\begin{equation}
 C=5\times \log\left(\frac{r_{80}}{r_{20}}\right),
\end{equation}

\noindent where $r_{80}$ and $r_{20}$ are the radii containing $80$ per cent and $20$ per cent of the galaxy's total light respectively \citep{Bershady00}. Elliptical galaxies are the most concentrated systems, and the concentration of stellar light decreases for later Hubble types \citep{Bershady00}. The quantitative values of C range roughly from $\mathrm{C}=2$ to $5$ with most systems with $\mathrm{C}>4$ being ellipticals, while disc galaxies have values between $3$ and $4$. The lowest values correspond to objects with low central surface brightness and low internal velocity dispersion \citep{Graham01,Conselice02}.

The \textbf{clumpiness} (or smoothness) describes the fraction of light in a galaxy that is contained in clumpy distributions. It is measured as follow:

\begin{equation}
 S=10\times\sum\frac{(I_0-I^{\sigma}_0)-B_0}{I_0},
\end{equation}

\noindent where $I^{\sigma}_0$ is the original image after reducing the resolution by a smoothing filter \citep{Bershady00,Conselice03b}. The sum is performed over the whole pixels of the image. The clumpiness (smoothness) is only measurable for the brightest and well resolved galaxies, thus it is of interest for our systems which are often selected by their clumpy light. By investigating this clumpy light with the S parameter we can get some idea of the origin of these structures and therefore of the galaxies themselves.

The clumpiness parameter S correlates with the star formation activity, as star-forming galaxies tend to have clumpy structures and, thus, high values of S \citep{Conselice03a}. The clumpiness of a galaxy's light compared with its asymmetry is also a good indicator for the origin of its structure as through a merger or another process dominated by star formation.

\section{Results}\label{sec.results}

\begin{figure*}
	\centering
	\includegraphics[width=0.82\linewidth]{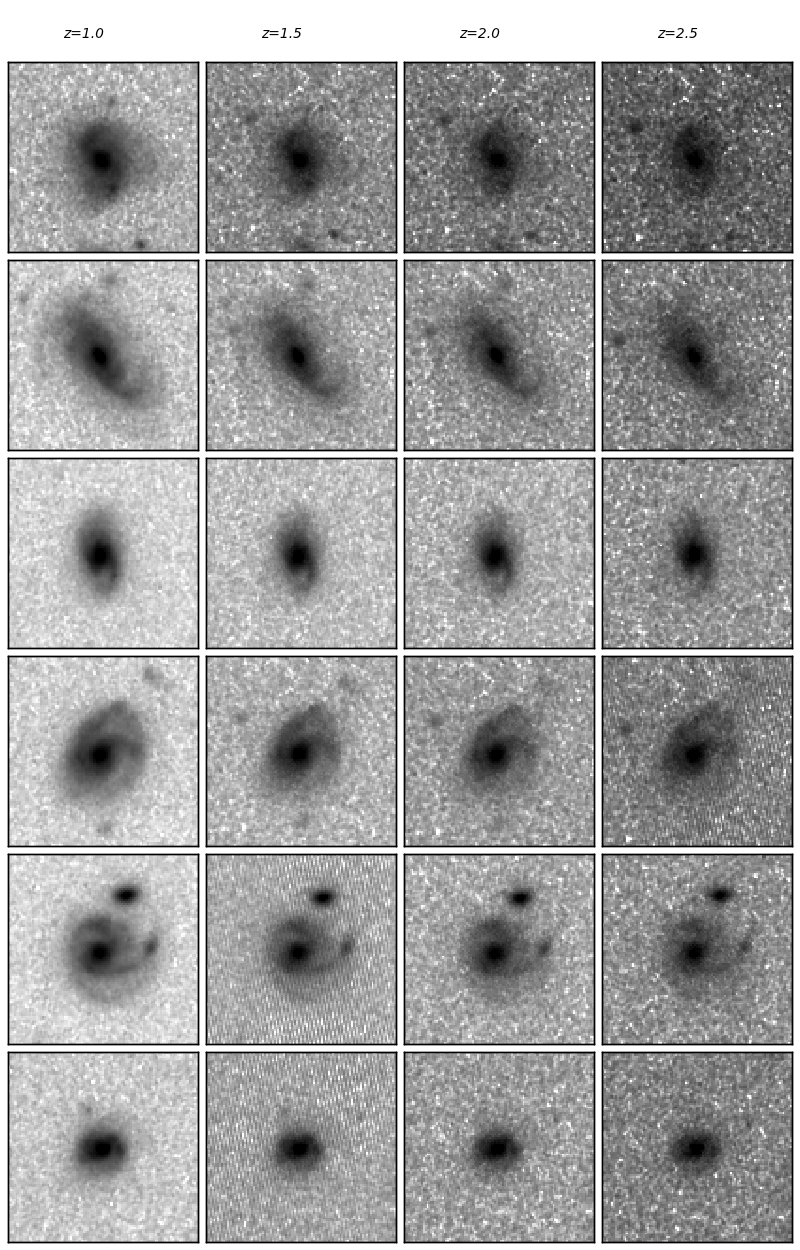}
	\caption{Artificially redshifted spiral galaxies in the CANDELS fields. Left panels show the original galaxies at redshift $z\sim1$. The next panels show the galaxies after being artificially redshifted at $z=1.5,2,2.5$, respectively. As can be seen most the spiral structures of these galaxies can still be distinguished within these images up to the highest redshifts we study them in. The postage stamps are $6 \times 6$ arcseconds in size.}\label{fig8}
\end{figure*}

In the results section we discuss the properties of our distant spiral galaxy sample which we defined in the previous section. These results include information about the stellar masses of these galaxies, their stellar content, their morphology and likely origin as well as basic information about the fraction of galaxies which are of a disc-like nature at progressively high redshifts. We first begin by investigating the spiral galaxy fraction and how it evolves with time and then characterise these systems before discussing their likely origin.

\subsection{Spiral Fractions}\label{sec.results.fractions}

We investigate in this section how the fractions of the different types of disc galaxies vary as a function of redshift. We also investigate the fraction of all galaxies that are discs and how both evolve with redshift. This will give us some indication of when spiral patterns have formed in galaxies. However, one caveat to this is that we are unable to be certain of all the morphological classifications we use, and how complete we are in our sample of spirals. However, we discuss later the several ways in which we address this issue. This is not as critical an issue later when we discuss the individual properties of the spiral galaxies themselves.

\begin{figure}
	\centering
	\includegraphics[width=1\linewidth]{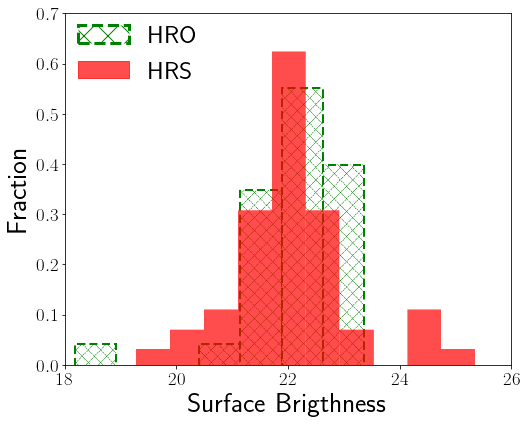}
	\caption{Surface brightness comparison between artificially redshifted galaxies at $z=2.5$ classified as spirals (high redshift simulated galaxies, HRS), and observed spiral galaxies at $z\sim2.5$ (high redshift original galaxies, HRO).}\label{fig9}
\end{figure}

\begin{figure}
	\centering
	\includegraphics[width=1\linewidth]{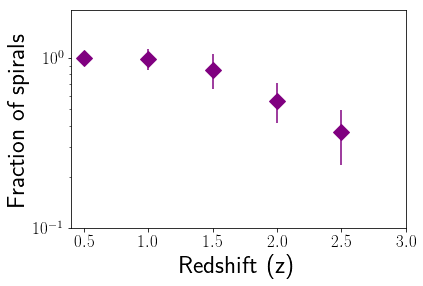}
	\caption{The fraction of recovered spiral galaxies after artificially redshifting systems at $z = 0.5$ to higher redshifts up to $z = 2.5$. This figure shows the fraction of spiral galaxies at $z=0.5$ that, after being artificially redshifted at different redshifts, are still able to be classified as spirals. For example, we select spiral galaxies at $z=0.5$ based on our methodology in \S 2 and artificially redshift them to how they look at $z=1$, and then calculate the fraction of these galaxies that are still classified as spirals. We also select the spiral galaxies at $z=1$ and after artificially redshifted them to $z=1.5,2,2.5$, we calculate the fraction at each redshift that still shows spiral patterns.}\label{fig10}
\end{figure}

We start by showing the evolution of the fraction of different disc and spiral types in Figure \ref{fig7}. What we find is that the fraction of disc galaxies classified as spiral discs increases with decreasing redshift by a factor of $14\pm3$. However, this could be due to redshift effects, which may cause spiral patterns to be more visible at lower redshifts. Galaxies classified as smooth discs (selected visually) have a similar fraction to spirals at the highest redshifts, but increase only by a factor of $\sim 7$ within our redshift range, while the fraction of clumpy spirals increases only by a factor of $\sim 2$. In total, we find $474\pm22$ spiral or clumpy spiral galaxies in our sample, with $152\pm12$ of them at $z>1$. When we compare the evolution of the fraction of all the discs from this work combined with \cite{Mortlock13} (their Figure 5), we find that our fraction is almost twice as high, although it follows the same decrease with increasing redshift. However, if we remove the smooth discs, our fractions are factor of $1.2\pm0.5$ smaller. 

These results are not surprising, and can be explained, on one hand, by the fact that when we include the smooths discs, we are likely including spheroids that have been missclassified as smooth discs. On the other hand, by our more restrictive selection of disc (when excluding the smooth discs) we are perhaps missing actual disc/spirals in that selection. We address this issue with simulations, which we discuss below.

A major issue with all of this observed evolution, and even prior studies on this \citep[e.g.,][]{Mortlock13}, is that many of the galaxies we would classify as a spiral would be hard to identify as such at higher redshifts due to the effects of redshift. Redshift creates a galaxy at lower signal noise and, at some redshifts, a lower resolution. To investigate to what extent the trends observed with redshift are due to these redshift effects, and not due to a real physical evolution, we have artificially redshifted all the galaxies at $z\sim 0.5$ in our sample to $z=1$, and from $z\sim1$ to higher redshifts ($z=1.5,2,2.5$), using the \textsc{ferengi} code from \cite{Barden08}. This procedure modifies the angular size and the surface brightness (dimming) due to cosmological effects, and also takes into account the brightness increase of high redshift objects from stellar population evolution \citep[see also][for a description of this]{Whitney20}. We therefore simulate how we would observe these same galaxies if they were at a higher redshift, following the same method as in \cite{Margalef16}. We then reclassify them following the decision tree shown in Figure \ref{fig1}, to ensure we classify the same way as Galaxy Zoo, by both BM-B and CC.

In Figure \ref{fig8} we show some examples of galaxies classified as spirals at $z\sim 1$ and how they appear at different redshifts, after artificially redshifting them. We then visually inspect this sample of galaxies and classify them as spirals (either `normal' spirals or clumpy spirals), or non spirals at each redshift. In Figure \ref{fig9} we show that the surface brightness distributions of galaxies classified as spirals artificially redshifted to $z=2.5$ (high redshift simulated galaxies, HRS) matches the real spirals observed at $z\sim2.5$ (high redshift original galaxies, HRO), which indicates that our artificially redshifted spiral galaxies are comparable to the real spirals at high redshift in terms of their surface brightness distributions. This is required, as if there were differences, it would imply that our artificially redshifted galaxies were either easier or harder to detect and analyse.

In Figure \ref{fig10} we plot the fraction of spirals that are still classified as such in the artificially redshifted galaxy images at different redshifts. It is clear from this figure that the ability to classify spirals decreases with increasing redshift and, in fact, it does so rapidly from $z=1$ to higher $z$. It is thus possible, or even likely, that we are missing some spirals at high redshift because the current instrumentation (WFC3) is not able to resolve spiral structures at $z>1$, meaning that some spirals may have been classified as smooth discs.

\begin{figure}
	\centering
	\includegraphics[width=1\linewidth]{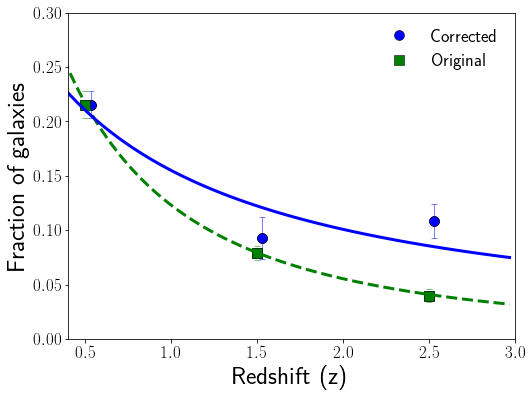}
	\caption{Fraction of spirals and clumpy spirals observed at different redshifts, taken as a fraction of all galaxies within our mass cut (green squares). This fraction increases to higher values after taking into account redshift effects (blue circles). The lines show the best fitting to power law functions.}\label{fig11}
\end{figure}

The observed evolution of the spiral fraction as a fraction of {\em all} galaxy types within the mass selection we use (Figure \ref{fig7}) is thus likely caused partially by redshift effects. After taking into account these effects, we observe that there is an increase in the number of galaxies classified as spirals or clumpy spirals from $z=1.5$ to $z=0.5$ (Figure \ref{fig11}). Thus at $z\sim2.5$ there may be at least twice as many spiral galaxies than what we observe. However, we still find a decline to higher redshifts even when taking this effect into account. 

We fit the fraction of disc and spiral galaxies as a fraction of all galaxy types as a function of redshift by a power-law function,

\begin{equation}\label{eq_power1}
f(z)=\alpha\cdot{}(1+z)^{\beta},
\end{equation}

\noindent where we fit $\alpha=0.48\pm0.08$ and $\beta=-2.0\pm0.1$ for the uncorrected fraction, and $\alpha=0.32\pm0.04$ and $\beta=-1.1\pm0.2$ for the corrected fraction. As shown in the best fit, the slope of the change is $\beta \sim -1$ for the fraction of disc galaxies, meaning that there is an increase in the relative abundance of discs at lower redshifts.

\begin{figure}
	\centering
	\includegraphics[width=1\linewidth]{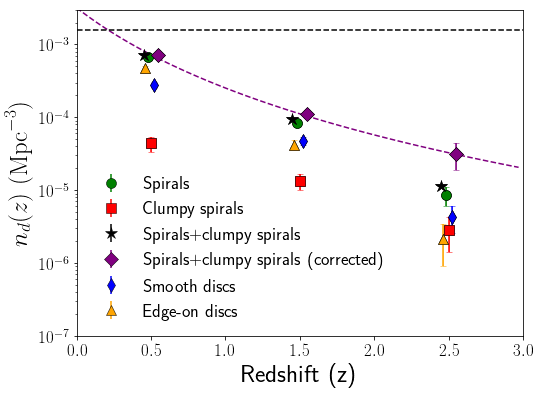}
	\caption{Number density evolution in units of co-moving volume Mpc$^{-3}$ for our different types of discs (for galaxies with $\log M_{\ast} >10$). The points are coloured as in Figure \ref{fig7}. The black stars show spirals and clumpy spirals together. The purple diamonds show the number density of spirals and clumpy spirals after correcting for redshift effects, and the dashed purple line shows the best fitting to a power-law function. The dashed black line is the number density of disc-dominated galaxies in the local universe. This is inferred from the fitted Schechter function for disc-dominated galaxies done by \protect\cite{Kelvin14} in the GAMA survey.}\label{fig12}
\end{figure}

In Figure \ref{fig12} we show the evolution of total number densities for the different types of spiral galaxies. Galaxies classified as either spirals or clumpy spirals increase in number density by a factor of $38\pm5$, from $z=2.5$ to $z=0.5$. However, if we consider the correction from the redshift effects, these spirals only rise by a factor of $14 \pm 2$ in number density, with the majority of the increase occurring from $z = 1.5$ to $z = 0.5$. Therefore, we observe in this figure a decrease in the number density of spiral galaxies at higher redshifts. The reason for this  decline is due to a few issues. This includes the fact that there is a smaller fraction of galaxies that are spiral, and there is a lower number density of galaxies over-all within our mass selection of M$_{*} > 10^{11}$ M$_{\odot}$.

The growth in number density of spiral galaxies plus clumpy spirals per co-moving volume in Mpc$^{3}$ units, after correcting for redshift effects, can be expressed as a power-law function:

\begin{equation}\label{eq_power2}
f(z)=\gamma\cdot{}(1+z)^{\delta},
\end{equation}

\noindent with fitted values $\gamma=0.0029\pm0.0003$ Mpc$^{-3}$ and $\delta=-3.5\pm0.3$ for the overall evolution of the number densities of disc galaxies. This line is plotted along with the number densities. 

\subsection{Stellar mass comparison}\label{sec.results.mass}

Next we investigate the difference in stellar mass for the four population of galaxies: spiral discs, clumpy spirals, smooth discs and edge-on discs. We are also interesting in the range of masses for these types compared to spirals and disc galaxies in the local universe. The reason for this is that there might be stellar mass thresholds whereby spiral arms develop or do not develop in galaxies at early times. Such a threshold would be a critical piece of information for understanding the physics behind the establishment of spiral arms in disc galaxies.

\begin{figure}
	\centering
	\includegraphics[width=1\linewidth]{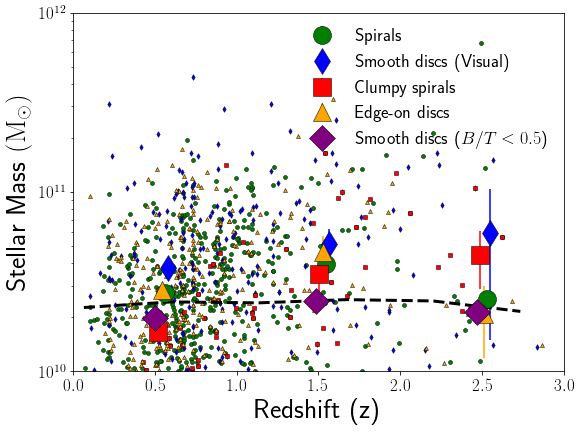}
	\caption{The stellar mass of our sample galaxies as a function of redshift for the different types of discs. Green circles are discs with a spiral pattern, red squares are discs with a clumpy spiral structure, yellow triangles are edge-on discs, blue diamonds are discs with no spiral pattern and purple diamonds are smooth galaxies with $B/T < 0.5$. The dashed line shows the average stellar mass of all the galaxies for the mass selected sample (including all types of discs, ellipticals and irregulars).}\label{fig13}
\end{figure}

In Figure \ref{fig13} we plot the stellar masses of the different types of discs as a function of redshift. On average, the smooth discs, visually classified, have the highest stellar masses, however, this may be due to a biased selection (the smooth discs visually classified are the most massive among the smooth discs). In fact, the average stellar mass of the smooth discs classified according to their $B/T$ is significantly lower. Taking this into consideration, the stellar masses of all the discs we study do not seem to differ much, on average, from each other, which may imply that stellar mass alone cannot explain the production of spiral patterns. There is also some evidence that the smooth discs have a higher mass at higher redshift. This likely due to resolution and signal to noise issues, and some of these galaxies would likely be clumpy spirals if seen at lower redshift. The number of smooth discs in the highest redshift bin is however quite small compared with those at lower redshift.

\begin{figure*}
  \centering
  \includegraphics[width=0.33\linewidth]{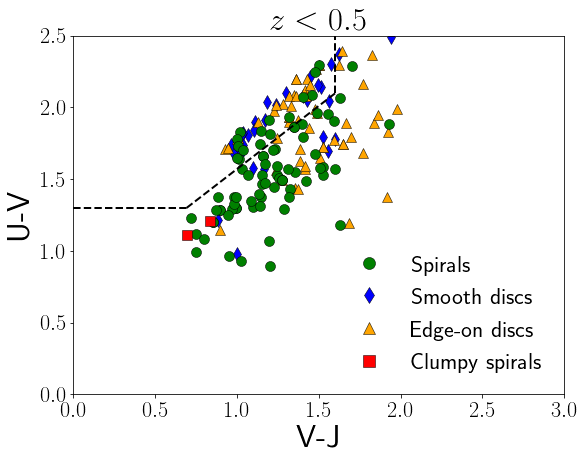}
  \includegraphics[width=0.33\linewidth]{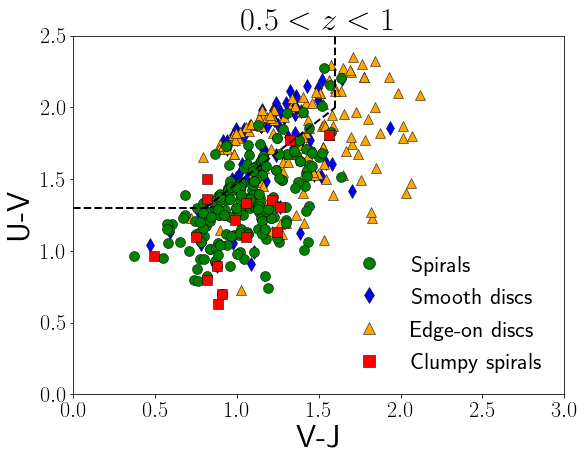}
  \includegraphics[width=0.33\linewidth]{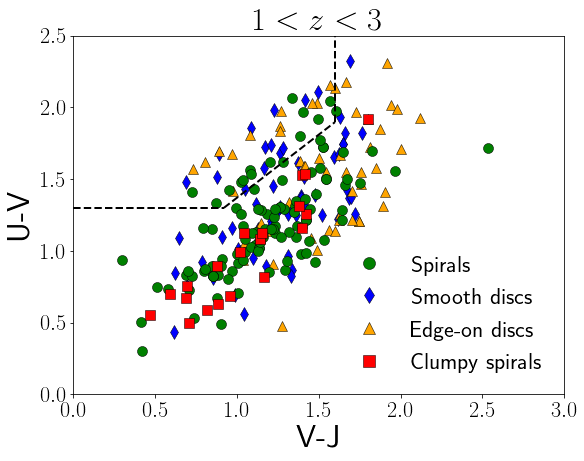}
  \caption{Rest-frame \textit{UVJ} diagrams in redshift bins from $0.2<z<3$. The left panel shows galaxies at $z<0.5$, the middle one galaxies at $0.5<z<1$, and the right panel for those at $1<z<3$. Green circles shows spiral discs, red squares, clumpy spirals; blue diamonds, smooth discs and yellow triangles, edge-on discs. The black dashed line shows the separation between passive and star-forming galaxies according to equation \eqref{ch4.eq.uvj}.}\label{fig14}
\end{figure*}

\subsection{\texorpdfstring{$UV\!J$}{UVJ} diagram. Dust vs. old stellar population}\label{sec.results.uvj}

Because we lack spectroscopy for these objects, we must use other aspects of the photometry to determine the properties of the spirals and disc galaxies in our sample. One of the ways this can be done is through the colours of the galaxies, and specifically by examining how the galaxies fall in colour-colour space which can be compared with models.

Using the CANDELS photometry and the \textsc{smpy} code \citep{Duncan14}, we compute the rest-frame \textit{U}, \textit{V} and \textit{J} Bessel magnitudes of our galaxies. We use the rest-frame $UV\!J$ colours to divide our sample in o red/passive and blue/star-forming systems, as this colour space allows for this separation. A galaxy is classified as red/passive if it satisfies the following criteria:

\begin{equation}\label{eq.uvj}
\begin{cases}
 (U-V)>1.3 \\
 (V-J)>1.6
\end{cases} 
\end{equation}

\noindent and the redshift dependent criteria
\begin{equation}
\begin{cases}
 (U-V)>0.88\times(V-J)+0.69\quad z<0.5 \\
 (U-V)>0.88\times(V-J)+0.59\quad 0.5<z<1.0 \\
 (U-V)>0.88\times(V-J)+0.49\quad z>1.0
\end{cases} \label{ch4.eq.uvj}
\end{equation}
\noindent and blue/star-forming otherwise.

Figure \ref{fig14} shows where our sample of galaxies lie in the \textit{UVJ} diagram. The majority of the spiral galaxies are in the star-forming region of the diagram, as might be expected. About $\sfrac{1}{3}$ of smooth discs and edge-one galaxies are in the passive region, but hardly any clumpy or spiral galaxies are found there. However, some of the galaxies classified as passive may be dusty star-forming galaxies. It is reassuring to confirm that visual classification of spirals and clumpy spirals, even at these high redshifts highly correlates with the star formation activity as observed in the \textit{UVJ} plane. In Table \ref{tab1} we summarize the fraction of passive and star-forming galaxies based on \textit{UVJ} definition.

\begin{table}
\centering
\begin{tabular}{lrr}
  & Star-forming & Passive \\ [0.5ex]
\hline \hline
\rule{0ex}{3ex}Spiral discs & $92 \pm 5\%$ & $7 \pm 1\%$  \\ 
\hline
\rule{0ex}{3ex}Clumpy spirals  & $97 \pm 13\%$ & $6 \pm 3\%$ \\
\hline
\rule{0ex}{3ex}Smooth discs &  $65 \pm 6\%$ & $32 \pm 4\%$  \\ 
\hline 
\rule{0ex}{3ex}Edge-on  discs &  $63 \pm 5\%$ & $37 \pm 4\%$ \\
\hline
\end{tabular}
\caption[Fraction of galaxies with different disc classifications that are passive and star-forming]{Fraction of galaxies with different disc classifications that are passive and star-forming at all redshifts, according to the $UV\!J$ selection from equation \eqref{eq.uvj}.}\label{tab1}
\end{table}

\begin{figure}
  \centering
  \includegraphics[width=1\linewidth]{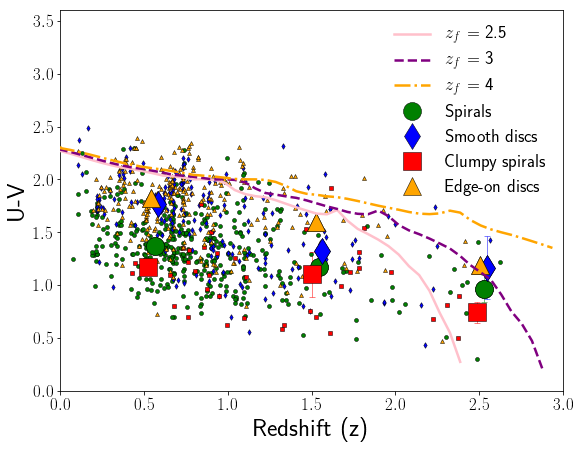}
  \caption{Rest-frame $U-V$ colour as a function of redshift. The points are coloured as in Figure \ref{fig14}. The lines shown represent $U-V$ colour evolution tracks derived from \protect\cite{Bruzual03} models with an initial formation at redshifts $z = 2.5, 3, 4$.}\label{fig15}
\end{figure}

We also examine the rest-frame colour $U-V$ as a function of redshift in Figure \ref{fig15}, which shows that spiral discs and clumpy spirals remain bluer than the rest of the disc at all redshift, while smooth discs and edge-on galaxies change more in colour and become redder as redshift decreases. We overplot $U-V$ colour evolution tracks derived from \cite{Bruzual03} models formed with a burst in star formation from a single stellar population at redshifts $z = 2.5, 3, 4$. None of the different types of discs is consistent with a single star formation burst, and in fact, they have bluer colours than the model tracks for a single burst, which implies that they must have undergone a more continuous star formation activity. This is something that we investigate later when examining the star formation histories of these objects. 

\subsection{Star formation activity}\label{sec.results.sf}

The star formation rate (SFR) is an important property to determine the formation state of these early discs/spirals, and to determine which galaxies contribute more at each epoch to the build up of stellar mass. In this work, SFRs are calculated by applying the \cite{Kennicutt98} conversion from the rest-frame \textit{UV} luminosity to SFR, and then correcting for dust using the \textit{UV} slope, following the same method as in \cite{Margalef18}.

The SFR before a dust correction is applied, and assuming a Chabier IMF, is expressed as follow:

\begin{equation}
 SFR(M_{\odot}\mathrm{yr}^{-1})=8.24\times10^{-29}L_{2800}(\mathrm{ergs} \ \mathrm{s}^{-1}\mathrm{HZ}^{-1}).
\end{equation}

\noindent Following the method described in \cite{Ownsworth16} we use ten \textit{UV} windows defined by \cite{calzetti94} to calculate the \textit{UV} slope, $\beta$, and then convert the slope into a dust correction using the \cite{Fischera05} dust model
$$A_{2800}=1.67\beta+3.71$$

As examined in detail in \cite{Ownsworth16} the dust measurements using the UV slope is consistent with that from measurements of dust extinction taken from SED fitting, as described in \S 2.3.  This is such that although we are recomputing this dust measurement, effectively the difference between the resulting SFRs in these methods is small and would not change our results.  We recalculate these with our direct measurements to be consistent with the methods in our previous work on this topic \citep[e.g,,][]{Margalef18}

\begin{figure*}
  \centering
  \includegraphics[width=0.49\linewidth]{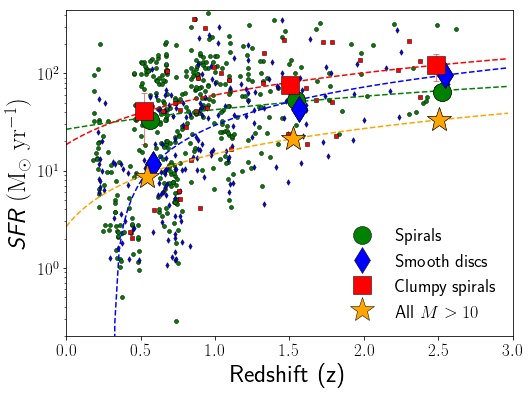}
  \includegraphics[width=0.49\linewidth]{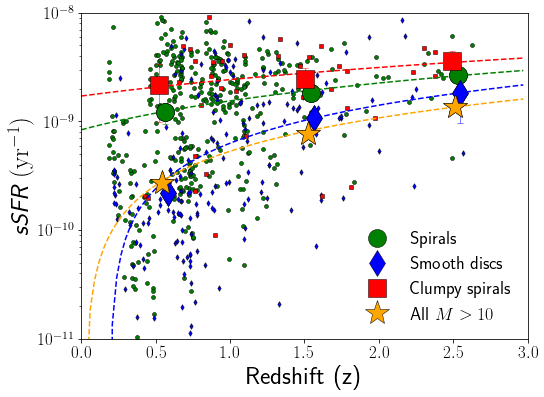}
  \caption{The SFR and sSFR (in log scale) as a function of redshift for the different types of discs we study in this paper. Green circles shows spiral discs, red squares show clumpy spirals and blue diamonds are smooth discs. The larger symbols are the median values at different redshift bins, and the error bars on these show the error on the median values. The big yellow stars show the median value of all the galaxies for the mass selected sample. The dotted lines are the best linear fit (in the linear fitting space) to the median values.}\label{fig16}
\end{figure*}

In Figure \ref{fig16} we show the SFR and specific SFR (sSFR) as a function of redshift. We include, for comparison, the average SFRs of all the galaxies with stellar masses $M>10^{10}\mathrm{M}_{\odot}$ as yellow stars on this plot. We have omitted the edge-on discs from these plots, as they present lower SFR values on average than the rest of the discs. This is likely an indications that the dust correction for the edge-on discs is underestimated. The dotted lines are the best linear fit to those median values ($\mathrm{SFR/sSFR} = \alpha*z+\beta$). Remarkably, we find that the spirals and clumpy spirals have a nearly constant sSFR which declines much slower at lower redshift than the smooth discs or the total mass selected sample.

The median value of SFR for spiral galaxies remains mostly constant with redshift at about $\mathrm{SFR}=40\pm 3\ M_{\odot}\mathrm{yr}^{-1}$ for spirals and $\mathrm{SFR}=50\pm 7\ M_{\odot}\mathrm{yr}^{-1}$ for clumpy spirals. We see that the spirals and clumpy spirals have the highest SFRs at all redshifts of all galaxy types, except at $z>1.5$, where the SFR of smooth discs is as high as for the spirals, however this could be caused by that fact that some of these galaxies could be spirals but, due to redshift depth/resolution effects, they are classified as smooth discs. The average SFR for smooth discs over all redshifts is $\mathrm{SFR}=19\pm43\ M_{\odot}\mathrm{yr}^{-1}$.

\begin{figure*}
  \centering
  \includegraphics[width=0.49\linewidth]{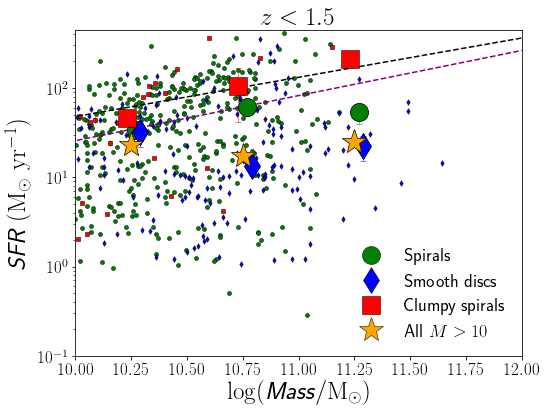}
  \includegraphics[width=0.49\linewidth]{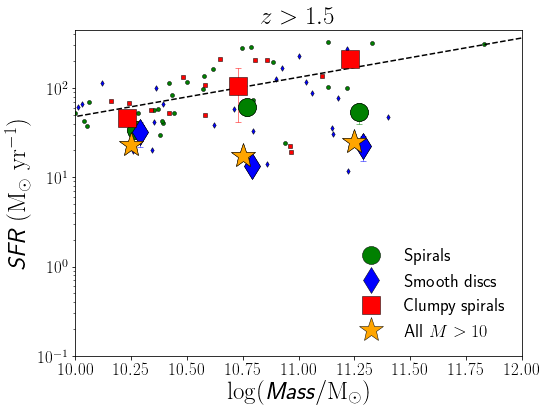}
  \caption{The SFR (in log scale) as a function of stellar mass for the different types of discs. The left panel shows galaxies at $z<1.5$ and the right panel galaxies at $z>1.5$. Points are coloured as in Figure \ref{fig16}. The big symbols are the median values at different stellar mass bins, and the error bars show the error on the median values. The dashed black line and purple line are the SFR-mass relation at $z=2$ and $z=1.5$, respectively, as found by \protect\cite{Whitaker12}.}\label{fig17}
\end{figure*}

A main-sequence of star-forming galaxies is observed at $z<1.5$ (left panel of Figure \ref{fig17}), and it appears to be formed mostly by spirals and clumpy spirals, and low mass smooth discs. Clumpy spirals appear to be above the main-sequence, as found in other studies, suggesting that they drive the forming phase of some high redshift galaxies. At $z>1.5$ the majority of the disc galaxies are consistent with the main-sequence of star-forming galaxies (right panel of Figure \ref{fig17}). Clumpy spirals dominate the SFR at all masses, with higher SFRs for higher stellar masses.

\subsection{Structure}\label{sec.results.structure}

\begin{figure*}
  \centering
  \includegraphics[width=0.48\linewidth]{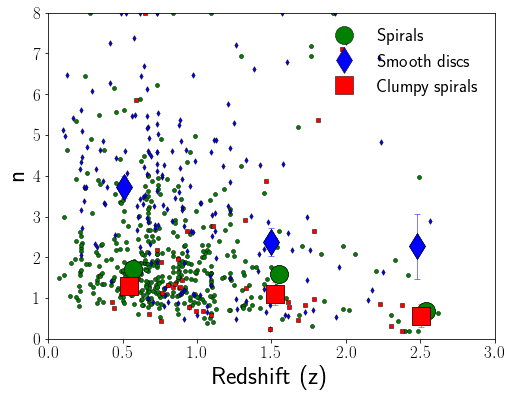}
  \includegraphics[width=0.49\linewidth]{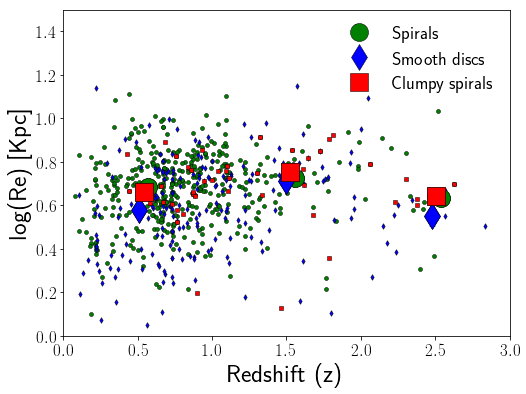}
  \caption{Evolution of the structural parameters: S\'ersic index (left) and effective radius (right), for the different types of discs we study in this paper. Points are coloured as in Figure \ref{fig14}. The large symbols are the median values at different redshift bins, and the error bars are the error on the median values. Note that we do not plot the edge-on discs, as the S\'ersic indices and effective radii do not describe them accurately.}\label{fig18}
\end{figure*}

To study the structural parameters of our sample of galaxies in the three CANDLES fields, we utilise the S\'ersic indices and effective radii measured by fitting S\'ersic profiles to the surface brightness of each galaxy  (see Section \ref{sec.data.structural}). Figure \ref{fig18} (left panel) shows that the S\'ersic index of spiral discs and clumpy spirals is lower at all redshifts than for smooth discs. This is unlikely due to redshift effects, as has been previously determined using similar resolution and depth data, e.g., \cite{Buitrago13}. This is still the case when including the smooth discs as selected by the ratio $B/T<0.5$. The spiral discs and clumpy spirals have on average $n<2.5$, although there is a slight increase in S\'ersic index $n$ with decreasing redshift. The S\'ersic index of the smooth discs also increases at lower redshift, and by $z=0.5$ they have on average $n>2.5$ ($n>2$ when including also the smooth discs selected by $B/T<0.5$). This may be a consequence of bulge formation.

The effective radii of our sample does not change significantly over cosmic time (right panel of Figure \ref{fig18}). All types of disc have on average similar sizes (although if we include the smooth discs selected by $B/T<0.5$, then the smooth discs are smaller than spiral objects at all redshifts). This is not the case for elliptical galaxies, which at redshifts $z\sim 2-3$ they are about $4-5$ times smaller than comparatively massive elliptical galaxies today \citep{Buitrago08}. Although the change in size for disc galaxies is less pronounced, it is established that disc galaxies also grow in size with decreasing redshift \citep{vanderWel12}. The absence of size evolution observed here is possibly due to redshift selection effects, as at high redshift the spirals classified by visual inspection will most likely be the biggest, while the spiral pattern of smaller galaxies may not be distinguishable at high redshift. 

In Figure \ref{fig19} we show the evolution of the CAS parameters for the different types of discs. The CAS parameters are computed in \cite{Mortlock15} for our disc sample. Overall, the concentration index, C, increases with decreasing redshift (left panel), which also reflects how the S\'ersic index $n$ evolves. We find that smooth discs have higher concentration indices than spirals at all redshifts. At $z>1$ we find $5$ spiral galaxies with very low concentration index $\mathrm{C}<2.2$, consistent with being Luminous Diffuse Objects \citep[LDOs,][]{Conselice04}. As expected, clumpy spirals have the highest asymmetry indices A (right panel), although spiral discs also have very high A values the more distant they are, which is an indication of spirals being more irregular at higher redshifts, as they settle into a normal disc morphology. 

We find that the smooth discs have slightly lower asymmetry values compared to other disc-like systems.  However we also observe that those classified as smooth discs at high redshift have a higher asymmetry compared with smooth discs at lower redshift. The reason for this is that the higher redshift smooth discs have a lower bulge to disc ratio, and thus more of their light is in a disc component.  This disc component can be lopsided or structurally asymmetric in bulk. As can be seen in Figure \ref{fig4}, these discs, whilst smooth, still have substructure in them.  Thus, the class of smooth discs is not homogeneous through all redshifts, but evolves with time such that the average bulge to disc ratio decreases at higher redshift. The clumpiness S (not plotted), decreases as well with lower redshift for all types of discs. In general we expect these values to become higher at higher redshifts once redshift effects are corrected for. Thus, the trend of a more clumpy and asymmetric morphology for spirals at higher redshifts is a strong result from our analysis.

\section{Conclusions}\label{sec.conclusions}

We have used the CANDELS fields to discover spirals and disc galaxies at redshifts $z > 1$ in the UDS, GOODS-S and COSMOS fields. Within this classification scheme, we classify spirals up to redshift $z\sim3$ and characterise their properties. One discovery we make is that some spiral galaxies at $z>2$ are likely misclassified as smooth discs due to redshift effects, as lower resolution at high redshift causes the spiral pattern to blend with the rest of the galaxy. We carry out simulations (by artificially redshifting galaxies) to correct for this effect. Our major findings in this paper including:

\begin{figure*}
  \centering
  \includegraphics[width=0.48\linewidth]{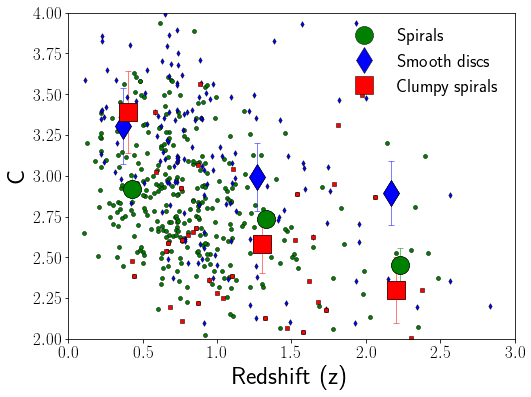}
  \includegraphics[width=0.48\linewidth]{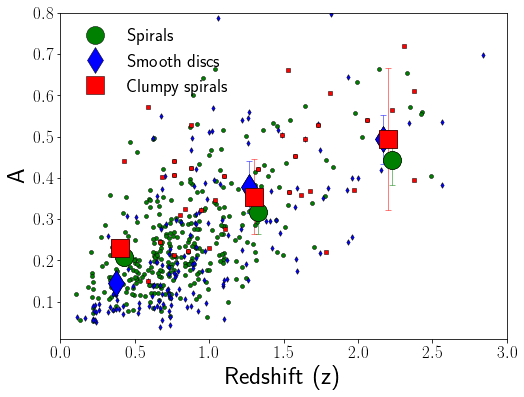}
  \caption{Evolution of the concentration C (left) and asymmetry A (right) values for our sample of disc galaxies. Points are coloured as in Figure \ref{fig14} with spirals, smooth discs and clumpy spirals plotted separately. The large symbols are the median values at different redshift bins for each of the types. We find that in general the spirals become more concentrated and less asymmetric as a function of time. }\label{fig19}
\end{figure*}

\noindent 1. In terms of number densities, spirals galaxies increase by a factor of $14\pm2$ at $\log M_{\ast} >10$ from the highest redshift bin ($2<z<3$) to the lowest ($0.2<z<1$), after correcting for redshift effects. The greatest increase occurs at $z\sim1.5$. Spiral patterns are thus formed very rapidly over a few Gyrs.

\noindent 2. Spirals at high redshift display different morphologies than in the local universe, with a much more clumpy appearance for typical systems. This likely the conduit method for their formation \citep[e.g.,][]{Guo15}.

\noindent 3. Spirals and clumpy spirals have low S\'ersic indices at all redshifts ($n<2.5$), which indicates a relatively flat surface brightness profile, linked with disky morphologies. Smooth discs (without spiral arms) become more concentrated (with higher S\'ersic index) as redshift decreases. On average both types have similar sizes that do not evolve with redshift. This however could be the result of redshift effects. We find that spiral patterns are more visible in larger galaxies and that therefore at the highest redshifts galaxies classified as spirals tend to be large in size. We cannot rule out that this is partially due to an observational bias.

The difference in structure between spirals and smooth discs is also observed in the CAS parameters. Spirals and clumpy spirals have lower concentration parameters than smooth discs, which may be an indication that smooth discs are more evolved as matter has been transferred from the outer regions to the inner ones due to secular evolution. We furthermore find that spirals and clumpy spirals have higher asymmetry values. In particular, clumpy spirals have on average the highest asymmetries and remain constant with redshift, while the asymmetry for spirals and smooth discs tend to be lower with decreasing redshift. The trends found for the concentration and asymmetry (Figure \ref{fig19}) are consistent with \cite{Bluck12}, who report a rise of asymmetry with redshift while the concentration decreases.

\noindent 4. The star formation rate of our identified spirals and clumpy spirals is the highest at all redshifts among the larger population of disc galaxies, and does not change significantly with redshift. Yet, there is a decrease in the star formation of spirals with redshift for the most massive galaxies ($\log M_{\ast}>11$). However, for spirals and clumpy spirals we find that the specific star formation remains high at all redshifts and does not decline at lower redshifts as the general population of massive galaxies does.

This trend with mass is also observed for the smooth discs. Massive galaxies shut down their star formation before lower mass galaxies for smooth discs and spiral galaxies. This is not the case however for clumpy spirals that dominate the SFR at all masses. 

At high redshift ($z>1.5$) the main-sequence of star-forming galaxies is made up of all types of discs, while at lower redshift it is formed mostly by spirals and clumpy spirals. This is in agreement with the position these galaxies occupy in the $UV\!J$ diagram. Spirals and clumpy spirals lie in the star-forming region of the diagram at all redshifts. Smooth discs are preferentially in the star-forming region as well, but about $32\pm 4$ per cent of these galaxies are in the passive region, especially at $z<1$.

Although it could be expected that spirals and clumpy spirals have the highest stellar masses, we find that on average all type of discs have similar stellar masses. Therefore, there is no trend with mass for producing spiral structure. Therefore within our mass limit we do not find that higher mass galaxies are more likely to contain a spiral structure.

Overall, we find that spiral galaxy formation occurs primarily within the redshift range of $1.5 < z < 3$ and sets in during a very active star formation mode in galaxies. There is still some evidence that a small fraction of spirals form at $z > 3$, outside the range of this study, which will have to be investigated with \textit{JWST}. Future observations of these objects with IFUs with a high spatial resolution will also help determine their origin and their kinematic and dark matter properties.

\section*{Acknowledgements}

We thank the CANDELS team for their support in making this Paper possible, as well as STFC in the form of a studentship, and the University of Nottingham for financial support.

\section*{Data availability} The data underlying this article are available and can be found in HST CANDELS archive.

\bibliographystyle{mnras}
\bibliography{references}
\clearpage

\bsp
\label{lastpage}
\end{document}